\pdfoutput=1
\documentclass{svmult}
\usepackage{graphicx}
\usepackage{amsmath}
\usepackage{amssymb}

\newcommand{\ttt}{\texttt}

\newcommand{\mca}{\mathcal}

\begin{document}

\title*{Constructions from Dots and Lines}
\author{Marko A. Rodriguez$^1$ \and Peter Neubauer$^2$}

\institute{
	AT\&T Interactive
	\texttt{markorodriguez@attinteractive.com}\\	
	\and
	NeoTechnology
	\texttt{peter.neubauer@neotechnology.com}
}

\maketitle

\footnotetext[1]{Rodriguez, M.A., Neubauer, P., ``Constructions from Dots and Lines," Bulletin of the American Society for Information Science and Technology, volume X, number X, pages X--X, ISSN:1550-8366, 2009.}

\begin{abstract}
A graph is a data structure composed of dots (i.e.~vertices) and lines (i.e.~edges). The dots and lines of a graph can be organized into intricate arrangements. The ability for a graph to denote objects and their relationships to one another allow for a surprisingly large number of things to be modeled as a graph. From the dependencies that link software packages to the wood beams that provide the framing to a house, most anything has a corresponding graph representation. However, just because it is possible to represent something as a graph does not necessarily mean that its graph representation will be useful. If a modeler can leverage the plethora of tools and algorithms that store and process graphs, then such a mapping is worthwhile. This article explores the world of graphs in computing and exposes situations in which graphical models are beneficial.
\end{abstract}

\section{The Bits and Pieces of the Dots and Lines}

\begin{figure*}[h!]
	\centering
		\includegraphics[width=0.625\textwidth]{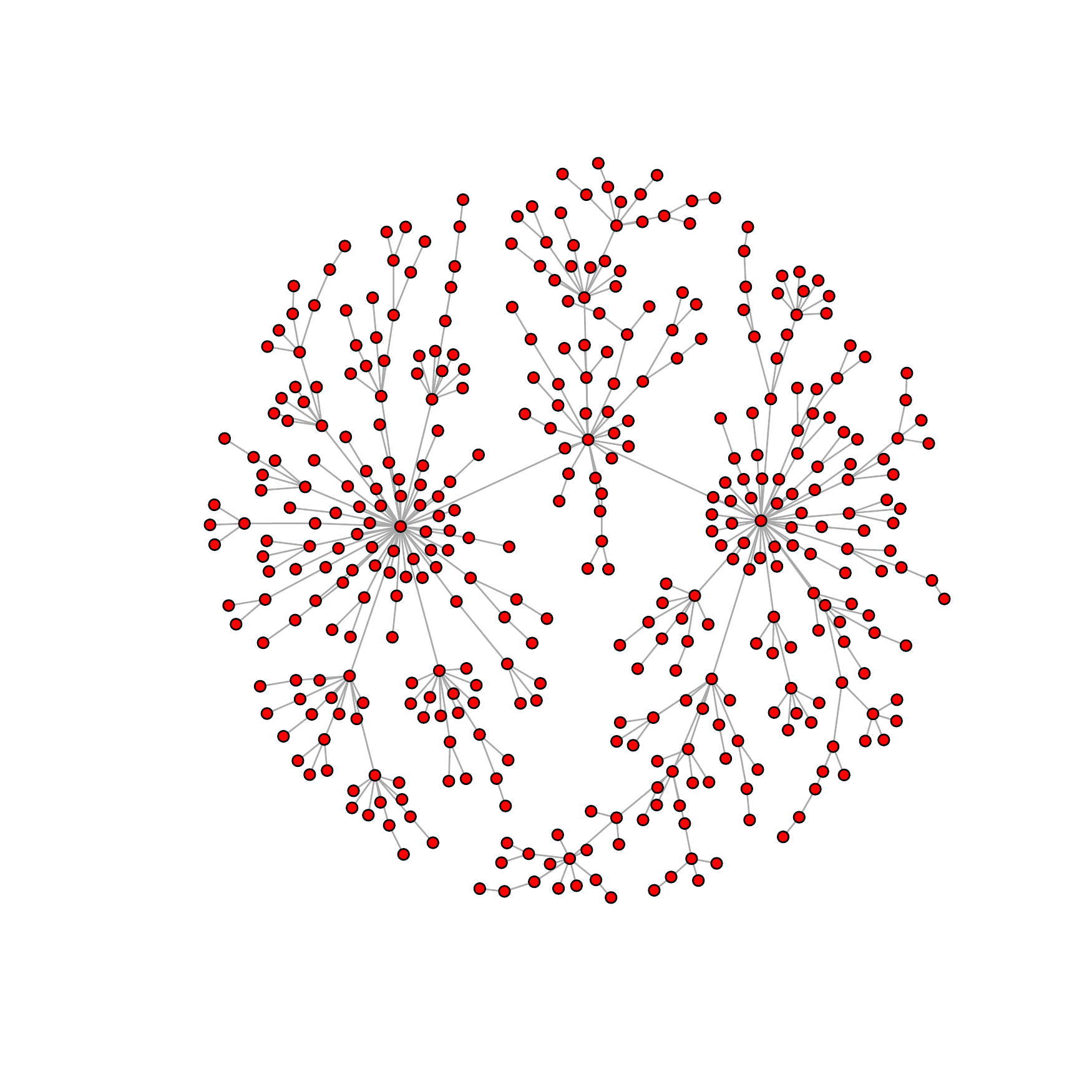}
	\caption{\label{fig:graph-example}The prototypical graph is the simple graph. In this structure there exists dots (i.e.~vertices) and lines (i.e.~edges). While the primitives are simple, their amalgamation can yield great complexity.}
\end{figure*}

A model is a representation of some aspect of reality. Many models can be thought of as a collection of objects (e.g.~people, concepts) and the relationships that exist between them (e.g.~friendships, subclasses). Such objects and relations form a network. Graphically, an object in a network can be denoted by a dot and a relationship can be denoted by a line. A structure formed by dots and lines is known as a graph---the mathematical term for a network \cite{intrograph:trudeau1976}. The most common type of graph is the simple graph. An example instance is diagrammed in Figure \ref{fig:graph-example}. In a simple graph there are a set of vertices (i.e.~dots) and a set of edges (i.e.~lines), where edges are undirected, connect two unique vertices (i.e.~no loops), and no two edges exist between the same pair of vertices.

\begin{figure*}[h!]
	\centering
		\includegraphics[width=0.95\textwidth]{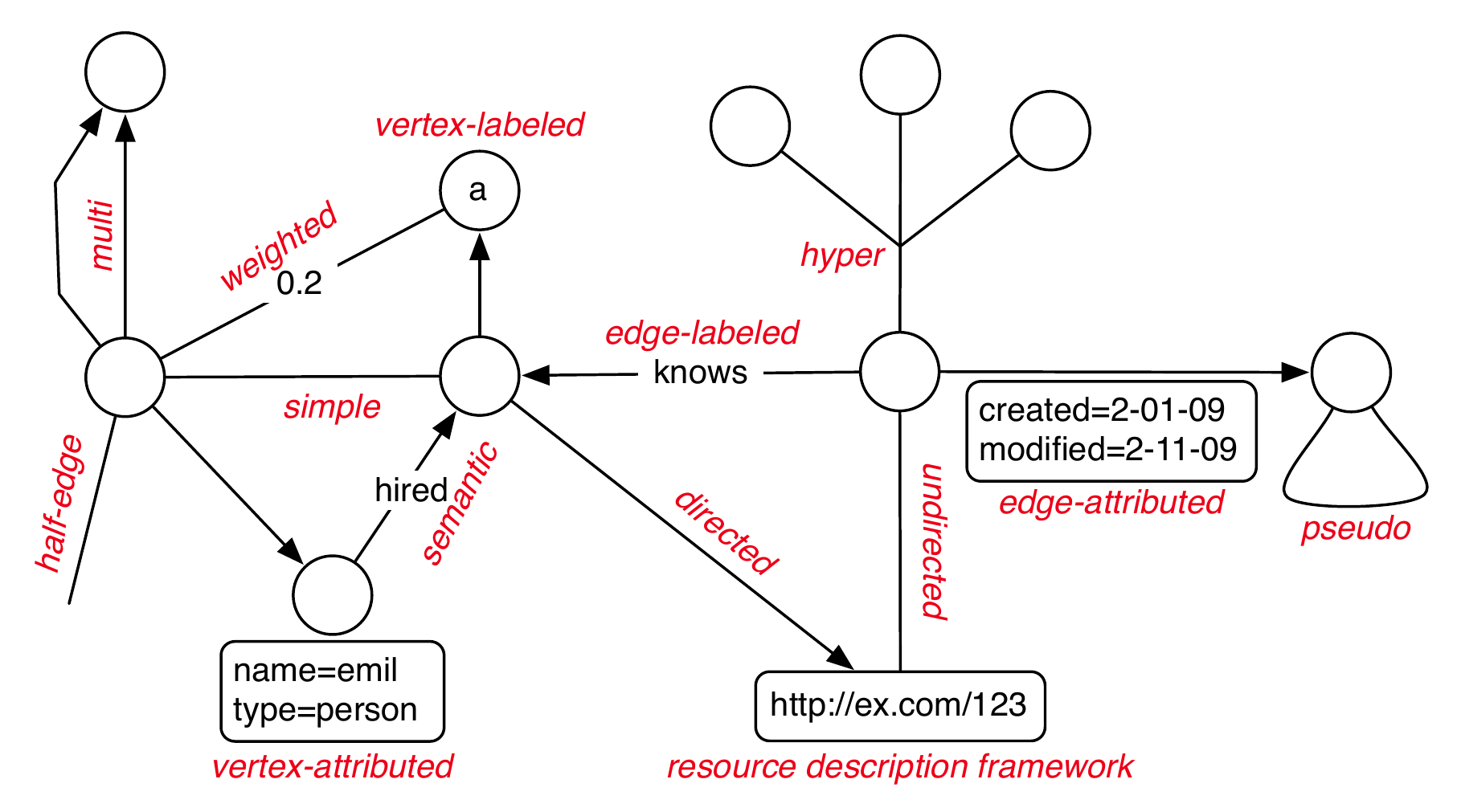}
	\caption{\label{fig:graph-types}There are numerous types of graphs. Many of the formalisms described can be mixed and matched in order to provide the modeler the expressivity necessary to capture the essential features of a domain.}
\end{figure*}

Contrary to the title of this article, dots and lines are not the only components in a graph modeler's toolkit. There are many more bits and pieces in the world of graphs. In practice, rarely are vertices and edges the only data contained within a graph. For instance, sometimes its useful to have a name associated with a vertex, a weight and direction associated with an edge, etc. From primitive dots and lines various bits and pieces can be added to yield a more flexible, more expressive graph. Figure \ref{fig:graph-types} diagrams a collection of different graph types. A short summary of each graph type is provided in the itemization below. Note that in many cases, these bits and pieces can be used in combination with one another (i.e.~they are not necessarily mutually exclusive).
\begin{itemize}
	\item \textbf{half-edge graph}: a unary edge (i.e.~an edge that ``connects" one vertex) has limited practical application and is primarily discussed in mathematics.
	\item \textbf{multi-graph}: there are many situations in which it is desirable to have multiple edges between the same two vertices.
	\item \textbf{simple graph}: the prototypical graph, where an edge connects two vertices and no loops are allowed.
	\item \textbf{weighted graph}: used to represent strength of ties or transition probabilities.
	\item \textbf{vertex-labeled graph}: most every graph makes use of labeled vertices (e.g.~an identifier).
	\item \textbf{semantic graph}: used to model cognitive structures such as the relationship between concepts and the instances of those concepts \cite{sowa:semantic1991}.
	\item \textbf{vertex-attributed}: used in applications where it is desirable to append non-relational metadata to a vertex.
	\item \textbf{edge-labeled graph}: used to denote the way in which two vertices are related (e.g.~friendships, kinships, etc.).
	\item \textbf{directed graph}: orders the vertices of an edge to denote edge orientation.
	\item \textbf{hypergraph}: generalizes a binary edge whereby an edge connects an arbitrary number of vertices \cite{hyper:gallo1993}.
	\item \textbf{undirected graph}: the typical graph that is used when the relationship is symmetric (e.g.~friendship).
	\item \textbf{resource description framework graph}: a graph standard developed by the the World Wide Web consortium that denotes vertices and edges by Uniform Resource Identifiers \cite{rdfintro:miller1998}.
	\item \textbf{edge-attributed graph}: used in applications where it is desirable to append non-relational metadata to an edge.
	\item \textbf{pseudo graph}: used to denote a reflexive relationship.
\end{itemize}
The list presented is not the complete space of all graph types, nor are the terms generally accepted in all domains. Many of these structures have been rediscovered in different domains and under different names. The important point is that there are numerous graph types and, consequently, there are systems and algorithms that exist to store and process them. 

\begin{figure*}[h!]
	\centering
		\includegraphics[width=0.85\textwidth]{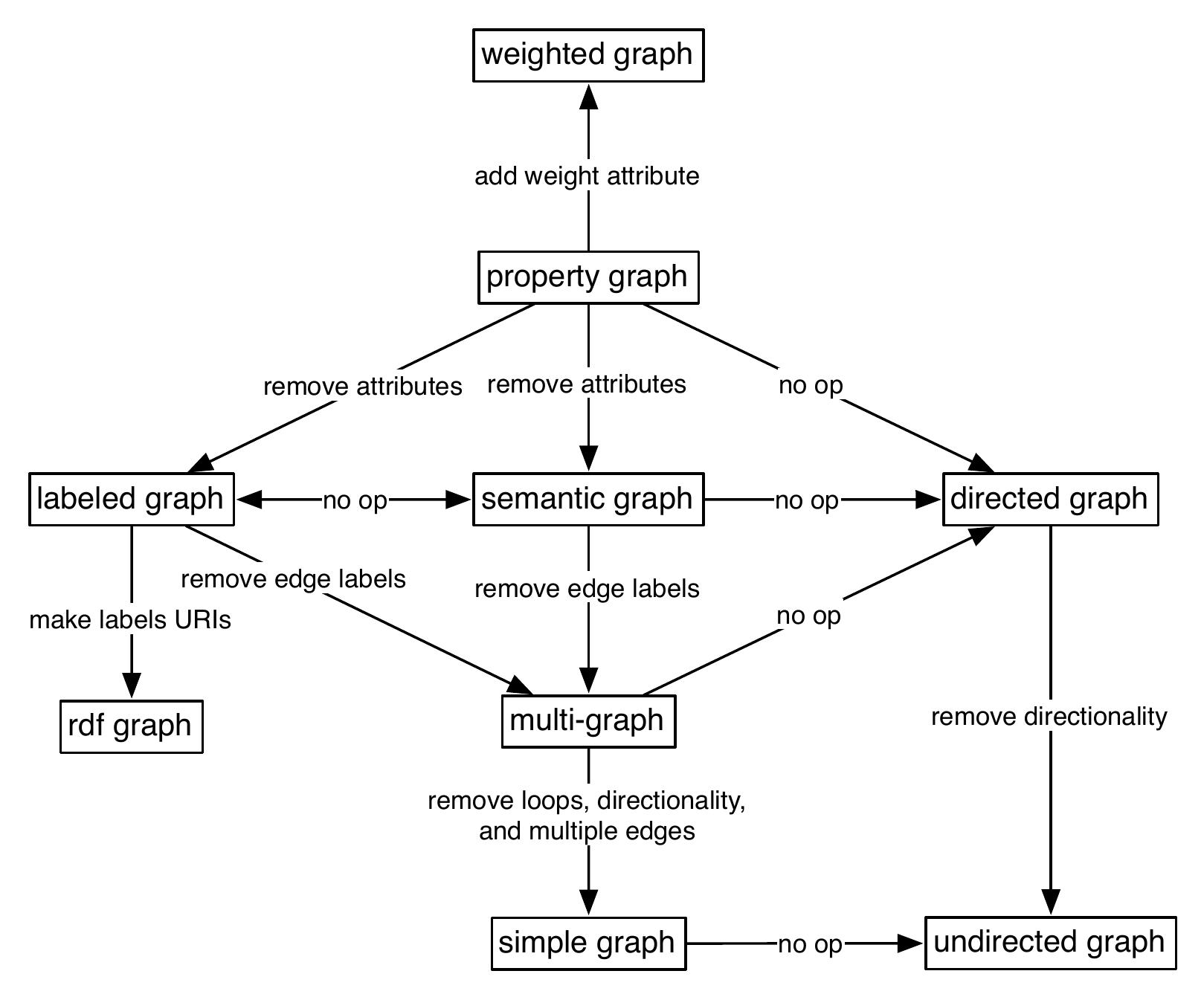}
	\caption{\label{fig:graph-types-morphisms}The property graph is a convenient structure because it contains most of the bits and pieces used in graph modeling. Simple morphisms of the the property graph yield other common graph structures. Thus, graph systems that support the property graph data model also, implicitly support other graph types.}
\end{figure*}

A common graph type supported by most graph systems is the directed, labeled, attributed, multi-graph---also known as a ``property graph." Graphs of this form allow for the representation of labeled vertices, labeled edges, and attribute metadata (i.e.~properties) for both vertices and edges. The property graph is common because by simply abandoning or adding particular bits and pieces, other graph types can be expressed. For example, by not allowing loops or multiple edges between two vertices, a simple graph is generated. By not allowing vertex/edge attributes, a standard semantic graph is generated. By restricting the vertex/edge labels to Uniform Resource Identifiers (URIs), a Resource Description Framework (RDF) graph is generated.\footnote{This is not completely true as an RDF graph makes use of URIs, literals, and blank/anonymous nodes. The distinction between these concepts are outside the scope of this article.}  By adding a weight attribute to an edge, a weighted graph is generated. The various graph types and the morphisms that yield one graph type from another are diagrammed in Figure \ref{fig:graph-types-morphisms}. Note the location of the property graph within this diagram. Finally, while it is possible to model a hypergraph in a property graph, it comes at the expense of using vertices in the property graph to denote both vertices and edges in the hypergraph. For this reason, there exist specialized hypergraph systems.\footnote{HyperGraphDB is an example hypergraph database that is available at: \ttt{http://www.kobrix.com/hgdb.jsp}.} For the remainder of this article, the more common property graph and its supporting technologies are discussed.

\section{Preserving Dots and Lines}

The computer science community has recently seen an explosion of database technologies. For decades, the relational database of Codd's relational algebra has been the primary storage and query mechanism for large data sets \cite{rdbms:codd1970}. However, with the continued growth of data and an increasingly variegated application landscape, new databases have emerged. In this space, no database is seen as the single solution to all problems. Instead, each database attempts to solve a particular data management issue. Itemized below is a short description of recent database types.
\begin{itemize}
	\item \textbf{document database}: These databases have the ``document" as their atomic entity. Such objects are semi-structured and usually represented in XML or JSON. A document can be retrieved by means of pattern matching a query document (i.e.~a semi-populated document) against all the documents contained in the database. The benefit of this model is that these databases scale horizontally with relative ease. This is due to the fact that documents lack references between one another. The drawback is that data is not interrelated and thus, cross database analyses are costly. For many web applications the document databases is a well-suited solution that supports data scale and a convenient symmetry between the document structure and the processing language (e.g.~languages that natively support XML and/or JSON). Examples of such databases include MongoDB\footnote{MongoDB is available at \ttt{http://www.mongodb.org/}.} and CouchDB\footnote{CouchDB is available at \ttt{http://couchdb.apache.org/}.}.
	\item \textbf{key/value store}: This family of databases is focused on the scaling of large amounts of data over a large number of machines and, in turn, supporting heavy read/write loads. Most of the databases in this class were inspired by Amazon's Dynamo \cite{dynamo:decandia2007}. A popular open-source key/value store is Tokyo Cabinet.\footnote{Tokyo Cabinet is available at \ttt{http://1978th.net/tokyocabinet/}.}
	\item \textbf{triple/quad store}: Triple/quad-stores were developed to support the demands of the Semantic Web/Web of Data/Linked Data community. These databases are optimized for storing and querying data represented according to the Resource Description Framework (RDF) \cite{rdfintro:miller1998}. Typical use cases include description logic reasoning \cite{baader:dl2003} and SPARQL-based graph pattern matching \cite{sparql:prud2004}. AllegroGraph is a high-performance quad-store with a large suite of extensions and features.\footnote{AllegroGraph is available at \ttt{http://www.franz.com/agraph/allegrograph/}.}
	\item \textbf{column store}: Most column stores are modeled after Google's BigTable database \cite{bigtable:chang2006}. A big table is a sparse, distributed, persistent multi-dimensional sorted map. The map is indexed by a row key, column key, and a time-stamp. Real-world services implemented with BigTable include GoogleAnalytics and GoogleEarth. Cassandra is a popular open-source column store.\footnote{Cassandra is available at \ttt{http://cassandra.apache.org/}.}
	\item \textbf{graph database}: Graph databases are optimized for the efficient processing of dense, interrelated datasets. In these databases, the atomic entity is the graph as a whole. The typical data model is the property graph. By supporting the interrelation of data, graph databases allow for fast traversals along the edges between vertices \cite{traversal:rodriguez2010}. A popular graph database of this form is Neo4j.\footnote{Neo4j is available at \ttt{http://neo4j.org/}.}
\end{itemize}

\begin{figure*}[h!]
	\centering
		\includegraphics[width=0.82\textwidth]{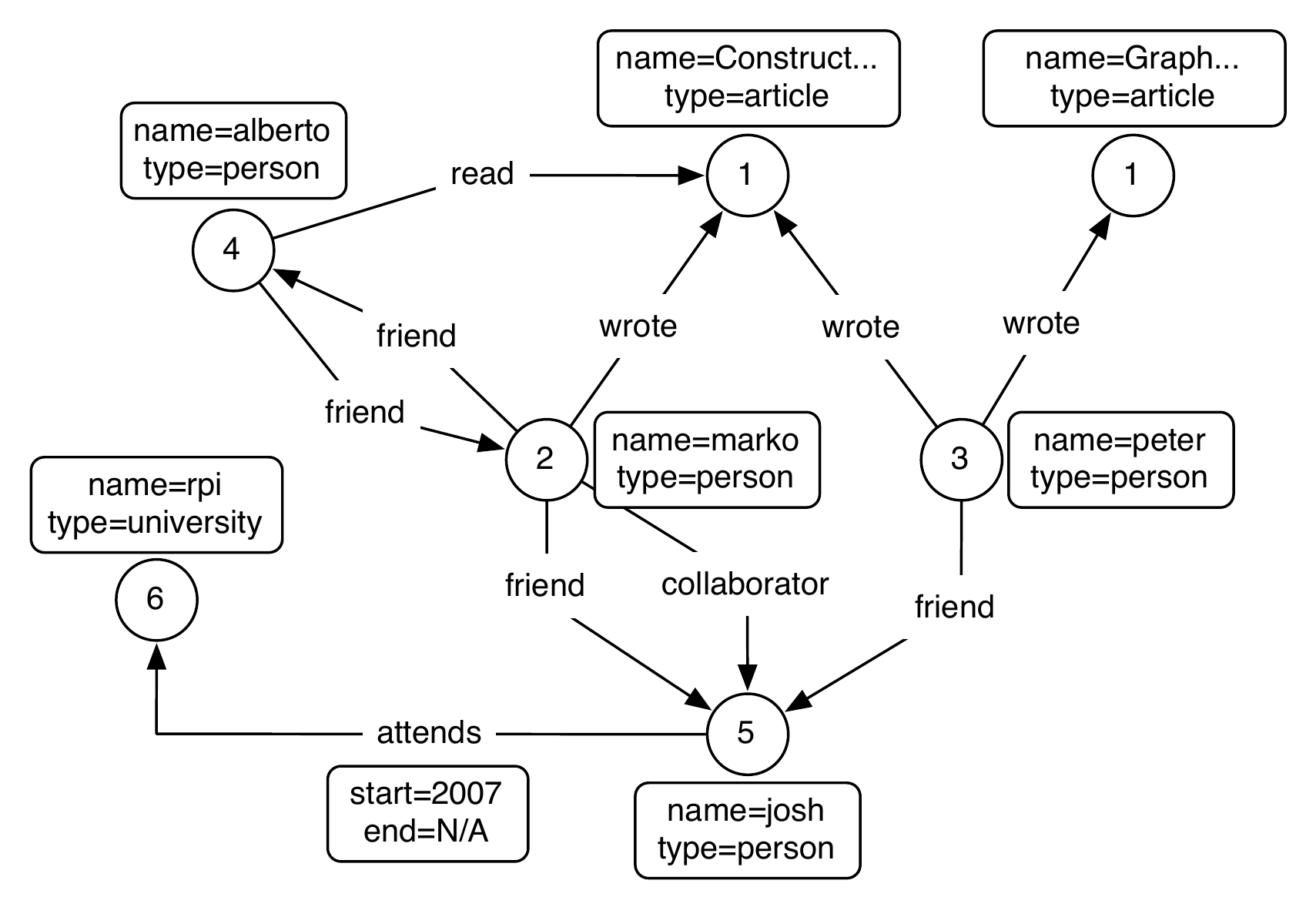}
	\caption{\label{fig:property-graph-example}A property graph is a directed, labeled, attributed, multi-graph. The edges are directed, vertices/edges are labeled, vertices/edges have associated key/value pair metadata (i.e.~properties), and there can be multiple edges between any two vertices.}
\end{figure*}

There are numerous databases in this growing space that were not mentioned. Moreover, there are other databases types not mentioned. It is out of the scope of this article to explore this space in depth. The interested reader is directed to related discussions, blog posts, and presentations that are made freely available on the Internet. Of particular relevance to this article is the graph database and the property graph data model. Figure \ref{fig:property-graph-example} diagrams a property graph containing people, their articles, and a university. In this particular domain model, each vertex has a \ttt{name} property and a \ttt{type} property. Edges denote both a directionality and a relationship type (i.e.~an edge label). Moreover, its possible to also include properties on an edge to further refine the way in which two vertices are related (e.g.~Josh started attending RPI in 2007).

\begin{figure*}[h!]
	\centering
		\includegraphics[width=1\textwidth]{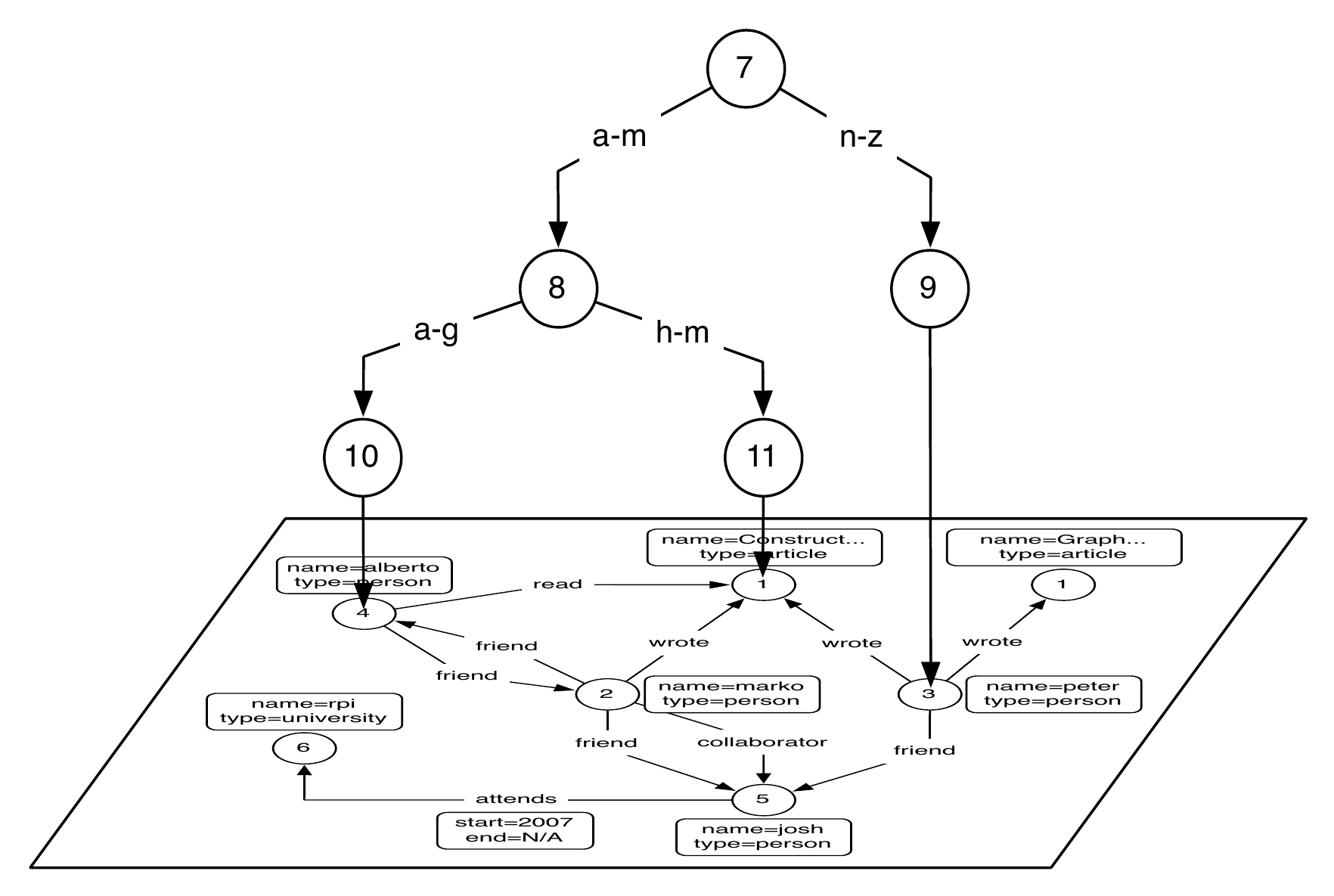}
	\caption{\label{fig:index-example}The index of the attributes/properties of the vertices and edges tend to be trees. A graph is a generalization of a tree. As such, graph databases allow for the modeling of the indices of the graph within the graph structure itself. For the sake of diagram clarity, the index does not touch every vertex with a \ttt{name} property. Finally, the edge labels of the index tree denote the ``bin'' that each sub-vertex is representing.}
\end{figure*}

A consequence of the flexibility of a graph is that other graph structures can be represented along with the domain model. A typical use case of such graph extensions include endogenous indices. An index is usually a tree-structure that allows for the fast look-up of elements within a collection. If there were no indices into a collection, then to determine if a particular element had a particular property, each element in the collection would have to be examined. The cost of a linear scan of this kind is $\mca{O}(n)$, where $n$ is the number of elements. What an index provides is the ability to partition the elements into increasingly fine-grained bins. Most indices have a lookup cost of $\mca{O}(\text{log}_2 \, n)$. While an index creates more data (the tree structure), it makes up for this cost by greatly increasing the speed of element retrieval. Figure \ref{fig:index-example} demonstrates a \ttt{name}-property index over the example graph diagrammed in Figure \ref{fig:property-graph-example}. Together, the domain model and the index of the domain model are seen as a single atomic entity. Searching for an element and moving between elements are accomplished by a unified framework: the graph traversal.

\section{Jumping from Dot to Dot}

The first aspect of using a graph is creating a graph. Once a graph has been created, it can be subjected to algorithms that quantify aspects of its structure, alter its structure, or solve-problems that are a function of its structure. At the root of any of these algorithms is the graph traversal \cite{traversal:rodriguez2010}. A graph traversal is a ``walk" along the elements of a graph---from vertex, to edge, to vertex, etc. As this walk proceeds, aspects of the graph can be saved or manipulated and in general, an algorithm can be computed. In principle, any of the data models and databases presented in the previous section (and including typical relational databases) can be used to represent and process a graph. However, when traversing a graph is the ultimate use case for a graph data set, then a graph databases is the optimal solution.\footnote{This is an import point. A graph database is optimized for graph traversals because elements (i.e.~vertices and edges) maintain direct references to their adjacent elements. It is this design choice that makes traversing a graph structure within a graph database fast and efficient.}

\begin{figure*}[h!]
	\centering
		\includegraphics[width=0.95\textwidth]{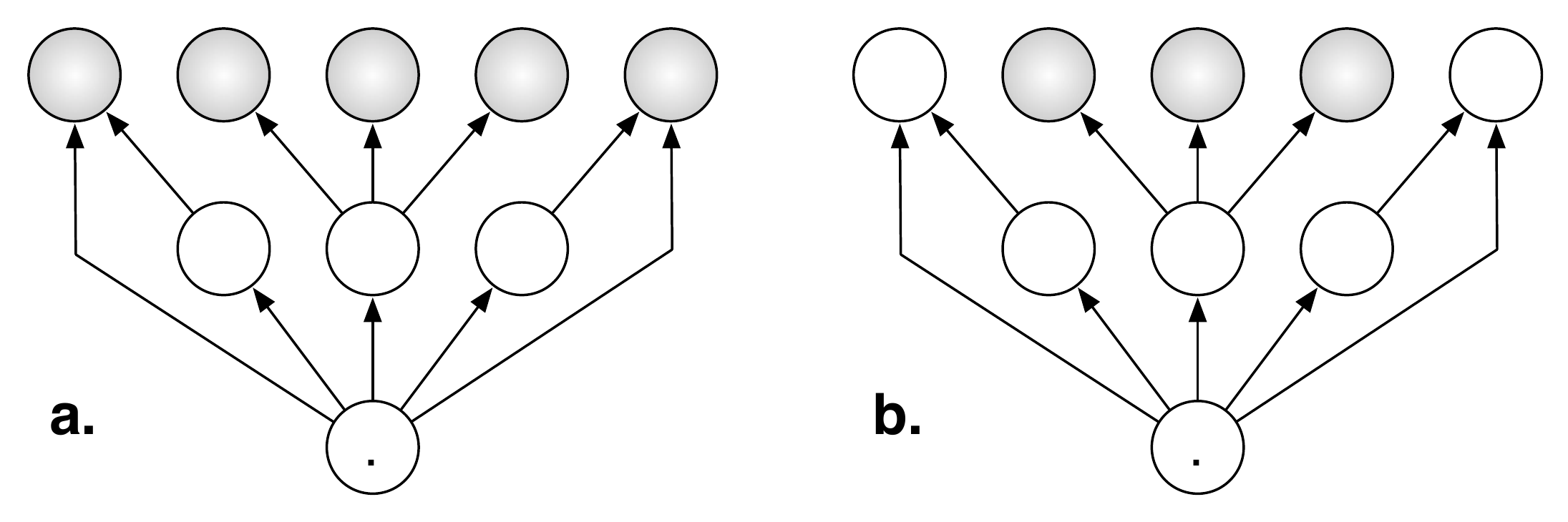}
	\caption{\label{fig:foaf-example}a.) The grey vertices denote the friends of the friends of the root vertex. b.) The grey vertices denote the friends of the friends of the root vertex who are not also the friends of the root vertex. For the sake of diagram clarity, the edges are not labeled. Assume that all edges are labeled ``friend."}
\end{figure*}

To get a better understanding of how graph traversals work, the examples in this section will be expressed in terms of a graph programming language called Gremlin.\footnote{Gremlin is available at \ttt{http://gremlin.tinkerpop.com/}.} In Gremlin, moving over vertices and edges is analogous, in many ways, to moving through the directory structure of a local filesystem. To demonstrate, a na\"ive friend-of-a-friend query is represented as follows:
\begin{verbatim}
	./outE[@label=`friend']/inV/outE[@label=`friend']/inV
\end{verbatim}
Reading from left to right, this expression states:
\begin{itemize}
	\item Start at the root vertex (\ttt{.}, i.e.~the vertex to evaluate the expression on).
	\item Traverse to all the outgoing edges of the root vertex (\ttt{/outE}).
	\item Filter out all edges that are not labeled ``friend" (\ttt{[@label=`friend']}).
	\item For all those friend-labeled edges, go to their incoming/head vertices (\ttt{/inV}).
	\item For all the friends of the root vertex, get their outgoing edges (\ttt{/outE}).
	\item Filter out all edges that are not labeled ``friend" (\ttt{[@label=`friend']}).
	\item For all those friend-labeled edges, go to their incoming/head vertices (\ttt{/inV}).
\end{itemize}
At the end of this expression, the resultant vertices are the friends of the friends of the root vertex. Figure \ref{fig:foaf-example}a diagrams the traversal, where the grey vertices are the returned vertices. This example is ``na\"ive" because in many cases, its important to retrieve the root vertex's friends of friends that are not also its friends. In such situations, the traverser must remember if a located friend-of-friend was not already a friend. In order to calculate the friend-of-a-friend, the friends must be determined first. Therefore, its possible to save this information for later use. This idea is diagrammed in Figure \ref{fig:foaf-example}b and the Gremlin expression is presented below, where the variable \ttt{\$x} references the friends of the root vertex.
\begin{verbatim}
	./outE[@label=`friend']/inV[g:assign(`$x')]/
	      outE[@label=`friend']/inV[g:except($x)]
\end{verbatim}

\begin{figure*}[h!]
	\centering
		\includegraphics[width=0.45\textwidth]{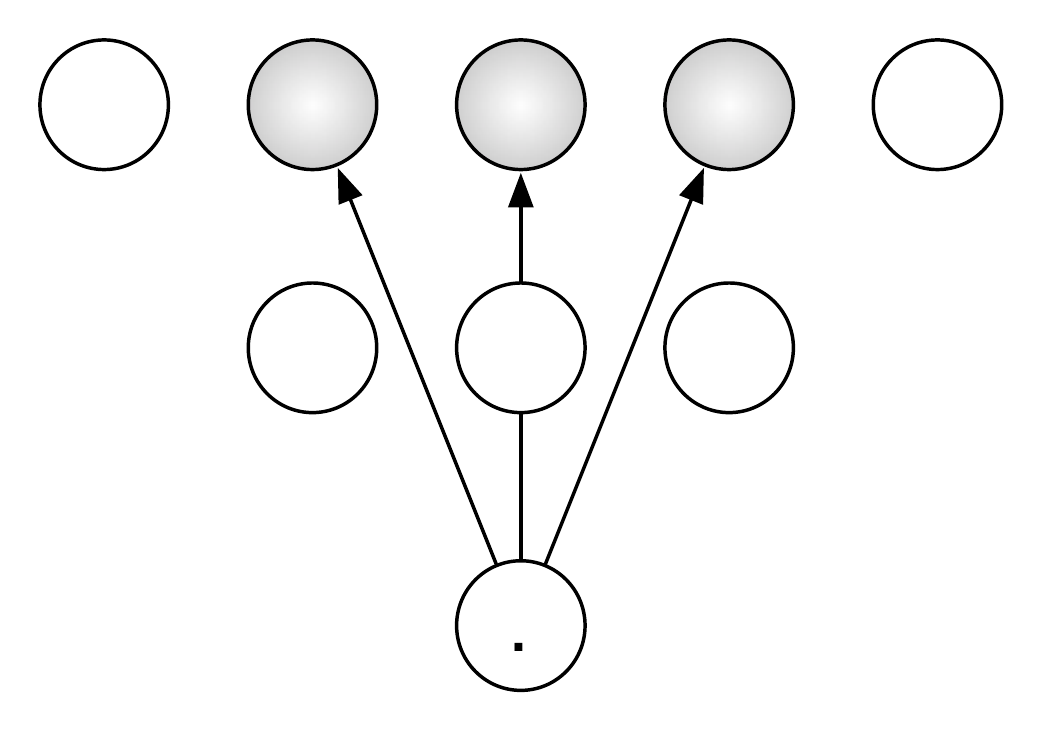}
	\caption{\label{fig:foaf-virtual-example}The evaluation of the friend-of-a-friend expression yields a path from the root vertex to the vertex's friends' friends. This path can be interpreted as a virtual/inferred/implicit/derived edge. For the sake of diagram clarity, no edges are labeled. Assume that all edges are labeled ``friend-of-a-friend."}
\end{figure*}

An important aspect of working with property graphs is that the edges are typed/labeled. The standard suite of graph algorithms found in most graph/network-theory textbooks are not immediately useful for property graphs \cite{netanal:brandes2005}. This is because, most graph algorithms have been developed for unlabeled graphs. When vertices can be related by many different ways and vertices can represent various types of objects, the meaning of the rankings, paths, etc. returned by standard graph algorithms are ambiguous. However, by interpreting a path through a graph as an edge, its possible to express standard graph algorithms on property graphs \cite{pathalg:rodriguez2009}. The previously presented Gremlin expression followed a path from the root vertex to its friends' friends. This path can be considered a ``virtual" (i.e.~inferred, derived) edge. From the perspective of this expression, a new implicit graph is created over the graph's vertices that only contains edges labeled ``friend-of-a-friend." This idea is diagrammed in Figure \ref{fig:foaf-virtual-example}. As such, this ``virtual" graph is equivalent to an unlabeled graph because all edges having the same meaning. Therefore, all the standard graph algorithms can be meaningfully applied to this derived graph---e.g.~the shortest path between person $A$ and person $B$ through their friends of friends. The benefit of edge-labeled graphs (e.g.~property graphs) is that there are as many types of rankings, scorings, etc. as there are types of paths that exist between the elements of the graph.

\section{Conclusion}

The concept of a graph was introduced in the late 19$^\text{th}$ century. During the many decades that followed, the world of graphs was primarily left to the toiling of mathematicians. In the last few decades, the sociology, physics, and computer science communities introduced a suite of algorithms and insightful realizations about the nature of graphs found in the real-world. Moreover, the increasingly large volume of data made available by the Internet has yielded datasets that reflect the graphs found in our technological and social systems. To satiate the need to handle and process these large-scale graphs, graph databases have come to the forefront. To make use of the graphs beyond simply representing their explicit structure, graph traversal frameworks and algorithms have been developed in order to shape graphs by driving the evolution of the entities that they model---e.g.~humans and their relationships to one another and the objects of their world \cite{faith2:rodriguez2009}.

\bibliography{../marko}

\begin{thebibliography}{10}

\bibitem{baader:dl2003}
Franz Baader, Diego Calvanese, Deborah~L. Mcguinness, Daniele Nardi, and
  Peter~F. Patel-Schneider, editors.
\newblock {\em The Description Logic Handbook: Theory, Implementation and
  Applications}.
\newblock Cambridge University Press, January 2003.

\bibitem{netanal:brandes2005}
Ulrick Brandes and Thomas Erlebach, editors.
\newblock {\em Network Analysis: Methodolgical Foundations}.
\newblock Springer, Berling, DE, 2005.

\bibitem{bigtable:chang2006}
Fay Chang, Jeffrey Dean, Sanjay Ghemawat, Wilson~C. Hsieh, Deborah~A. Wallach,
  Mike Burrows, Tushar Chandra, Andrew Fikes, and Robert~E. Gruber.
\newblock Bigtable: A distributed storage system for structured data.
\newblock In {\em {Proceedings of the 7th USENIX Symposium on Operating Systems
  Design and Implementation}}, pages 205--218, Berkeley, CA, 2006. USENIX
  Association.

\bibitem{rdbms:codd1970}
Edgar~F. Codd.
\newblock A relational model of data for large shared data banks.
\newblock {\em Communications of the {ACM}}, 13(6):377--387, 1970.

\bibitem{dynamo:decandia2007}
Giuseppe DeCandia, Deniz Hastorun, Madan Jampani, Gunavardhan Kakulapati,
  Avinash Lakshman, Alex Pilchin, Swaminathan Sivasubramanian, Peter Vosshall,
  and Werner Vogels.
\newblock Dynamo: {A}mazon's highly available key-value store.
\newblock {\em SIGOPS Operating Systems Review}, 41(6):205--220, 2007.

\bibitem{hyper:gallo1993}
Giorgio Gallo, Giustino Longo, Stefano Pallottino, and Sang Nguyen.
\newblock Directed hypergraphs and applications.
\newblock {\em Discrete Applied Mathematics}, 42(2-3):177--201, 1993.

\bibitem{rdfintro:miller1998}
Eric Miller.
\newblock An introduction to the {R}esource {D}escription {F}ramework.
\newblock {\em Bulletin of the American Society for Information Science and
  Technology}, 25(1):15--19, November 1998.

\bibitem{sparql:prud2004}
Eric Prud'hommeaux and Andy Seaborne.
\newblock {SPARQL} query language for {RDF}.
\newblock Technical report, World Wide Web Consortium, October 2004.

\bibitem{traversal:rodriguez2010}
Marko~A. Rodriguez and Peter Neubauer.
\newblock The graph traversal pattern.
\newblock [book chapter in review] arXiv:1004.1001, AT\&Ti and NeoTechnology,
  2010.

\bibitem{pathalg:rodriguez2009}
Marko~A. Rodriguez and Joshua Shinavier.
\newblock Exposing multi-relational networks to single-relational network
  analysis algorithms.
\newblock {\em Journal of Informetrics}, 4(1):29--41, 2009.

\bibitem{faith2:rodriguez2009}
Marko~A. Rodriguez and Jennifer~H. Watkins.
\newblock Faith in the algorithm, part 2: Computational eudaemonics.
\newblock In Juan~D. Vel{\'a}squez, Robert~J. Howlett, and Lakhmi~C. Jain,
  editors, {\em {Proceedings of the International Conference on Knowledge-Based
  and Intelligent Information \& Engineering Systems}}, volume 5712 of {\em
  Lecture Notes in Artificial Intelligence}, pages 813--820. Springer-Verlag,
  2009.

\bibitem{sowa:semantic1991}
John~F. Sowa.
\newblock {\em Principles of Semantic Networks: Explorations in the
  Representation of Knowledge}.
\newblock Morgan Kaufmann, San Mateo, CA, 1991.

\bibitem{intrograph:trudeau1976}
Richard~J. Trudeau.
\newblock {\em Dots and Lines}.
\newblock Kent State University Press, 1976.

\end{thebibliography}
\bibliographystyle{plain}

\end{document}